\documentclass{article}

\usepackage[utf8]{inputenc}		%
\usepackage[T1]{fontenc}		%
\usepackage{graphicx}			%
\usepackage{amssymb} 			% Symbols
\usepackage{amsmath} 			% Symbols
\usepackage{wasysym}
\usepackage{url}
\usepackage{fullpage}
\begin{document}

\title{A Physical Framework for the Earth System in the Anthropocene: Towards an Accountancy System}

\author{O. Bertolami\footnote{orfeu.bertolami@fc.up.pt}, F. Francisco\footnote{frederico.francisco@fc.up.pt}}

%\address{Departamento de Física e Astronomia and Centro de Física do Porto, Faculdade de Ciências, Universidade do Porto, Rua do Campo Alegre 687, 4169-007 Porto, Portugal}

\date{\small{Departamento de Física e Astronomia and Centro de Física do Porto,\\ Faculdade de Ciências, Universidade do Porto,\\ Rua do Campo Alegre 687, 4169-007 Porto, Portugal}}

\maketitle

\begin{abstract}
At a time when humanity has achieved global dominance at a scale that was previously thought impossible, it might also face an existential threat due to the consequences of that overwhelming influence on our common home, the Earth System (ES). In this work we explore how Physics may help us to understand the transitions that the ES is going through and lead us to a physically motivated accounting system that allows for setting boundaries to our negative influence on the ecosystems.
\end{abstract}

%-----------------------------------------------------------------------------%
%-- Body ---------------------------------------------------------------------%
%-----------------------------------------------------------------------------%

\section{Humanity in the Anthropocene \cite{Bertolami:2015,Bertolami:2018a,Bertolami:2018b}}

The world we inhabit in the early 21th century is characterized by global influences and tendencies. Cultural, economic, financial, sociological and scientific developments quickly spread across the world as information and transactions flow at velocities close to their physical limit.

Decisions are constantly being made under the pressure for quick answers. The capability to respond swiftly is often the very measure of their quality. By contrast, in today's world, urgency is not proportional to the risk that humans and the environment are exposed to, but rather to the rate of profit that can be collected for a given financial effort. This prevailing priority shifts the rationality of the decisions to time scales that are not compatible with the long-term well being of the majority and that might even threaten the very existence of human societies.

The examples are often disturbing and surprising. Important scientific discoveries, like nuclear energy for peaceful purposes or the widespread use of CFC compounds in the thermodynamic cycles of refrigerators and air conditioners were well meant on their roots. However, they both gave rise to unforeseeable environmental implications menacing humankind. Financial practices with no correspondence on the exchange of goods and services in the economy have led to systemic crises, global recession, unemployment, political turmoil and widespread suffering.

At the risk of oversimplifying, and from a rather euro-centric perspective, we can glance through what took place during the last few centuries that most directed influenced our time.

The cultural background embodied by the city-states Athens, Jerusalem and Rome gave origin, from the 14th century till the early 17th century, to the Renaissance. The Renaissance, the European expansion and the Mercantilism gave rise in the 17th century to the Scientific Revolution in Europe. The Scientific Revolution, the French Revolution and the British Industrial Revolution engendered fundamental changes that lead to the contradictions of what some historians called the “long” 19th century, from 1789 to 1914. These contradictions, through massive industrialization that was made possible by the knowledge of classical physics and chemistry and the abundance of raw materials arriving from colonies outside Europe, together with the political tensions among the capitalist world, fascism and soviet socialism were unresolved and led to two deadly World Wars. Europe was devastated twice and sentenced to a steady decline in its influence from 1945 onwards. The emerging world powers formed blocks around different economic and ideological conceptions, multinational capitalism versus Soviet central-command economies, and a tense balance was maintained under the threat of a nuclear war. In the early 1990s, the collapse of the Soviet Union and the widespread use of quantum-physics-based technology and the Internet reshaped the world once again.

We live now in a world intertwined by transport and communication networks that support economic, cultural, and financial transactions that involve the entire planet. A fast-moving world where the European Union, a unique experiment in supra-national governance, has been faltering, and China has emerged as a major key global player. The ensuing economic, social and political transformations will have implications in the life of billions of people.

The physical environment is sometimes seen simply as the setting where history takes place, but the world has now entered the Anthropocene. Even if the geological and stratigraphic debate is not yet fully settled, it is unquestionable that humans are the single strongest driving force of global environmental change. Especially, from 1950s onwards, there has been a great acceleration of the changes on the crust of Earth driven predominantly by the human intervention.

How can we balance these tendencies? How can we empower agents and institutions in the design and in the implementation of solutions?

In the next few decades, important decisions will have to be considered in order to halt the degradation of the environment and the steady destruction of ecosystems. It is foreseeable that without decisive measures to control the use of the natural resources, developments in quantum computing, artificial intelligence and genomic edition will boost a new robotic driven industry and economy, leading to major ecological disasters, massive unemployment and generalized poverty. The world population is expected to reach 9 thousand million people in 2050, more and more concentrated around China, India and Nigeria. Thus, in order to feed the world a considerable injection of resources will be required which will involve an increase in the area of cultivated land and of human labour while diminishing the ensued impact on the environment.

We suggest, likewise Friedrich Engels in his “Socialism: Utopian and Scientific” of 1880, that Utopia, as discussed by Thomas Moore in 1516 and many others after him, should be reconsidered in terms of the identifiable underlying economic mechanisms and that the 21st century scientific knowledge that allow for creating a sustainable and humane new world order, the “Scientific Utopia”. In our view, this methodological change is only achievable if all components of the so-called Earth System are internalized into a global and all-encompassing economic cycle. This new economic chain should also include all disruptive production processes and technologies that jeopardize the social cohesion so that the damage they cause are paid back to society. Actually, it should be remembered that the issue of development and economic performance involves a set of conditions such as political freedom, social opportunities, transparency guarantees, and protective security that can only be ensured by a collective ethical choice based on a principle of maximization of the common interest as defended by economists such as Amartya Sen \cite{Sen:2012}.

But how can we achieve the internalization of intangible goods and services provided by all the ecosystems?

Taxation and incorporation into the final price to consumers is an obvious solution for manufacturing and distribution of goods that harm the environment, cause unemployment and upset social harmony. But of course, these ideas can only work if set up on a global scale so that the “damage” cannot be exported. Clearly, these issues require not only profound political and social reorganization of the market economy, but also demand deep changes in the very principles of manufacturing and delivery of goods and an urgent implementation and intensification of the principles of a circular economy.

\begin{figure}
	\centering
	\includegraphics[width=0.7\columnwidth]{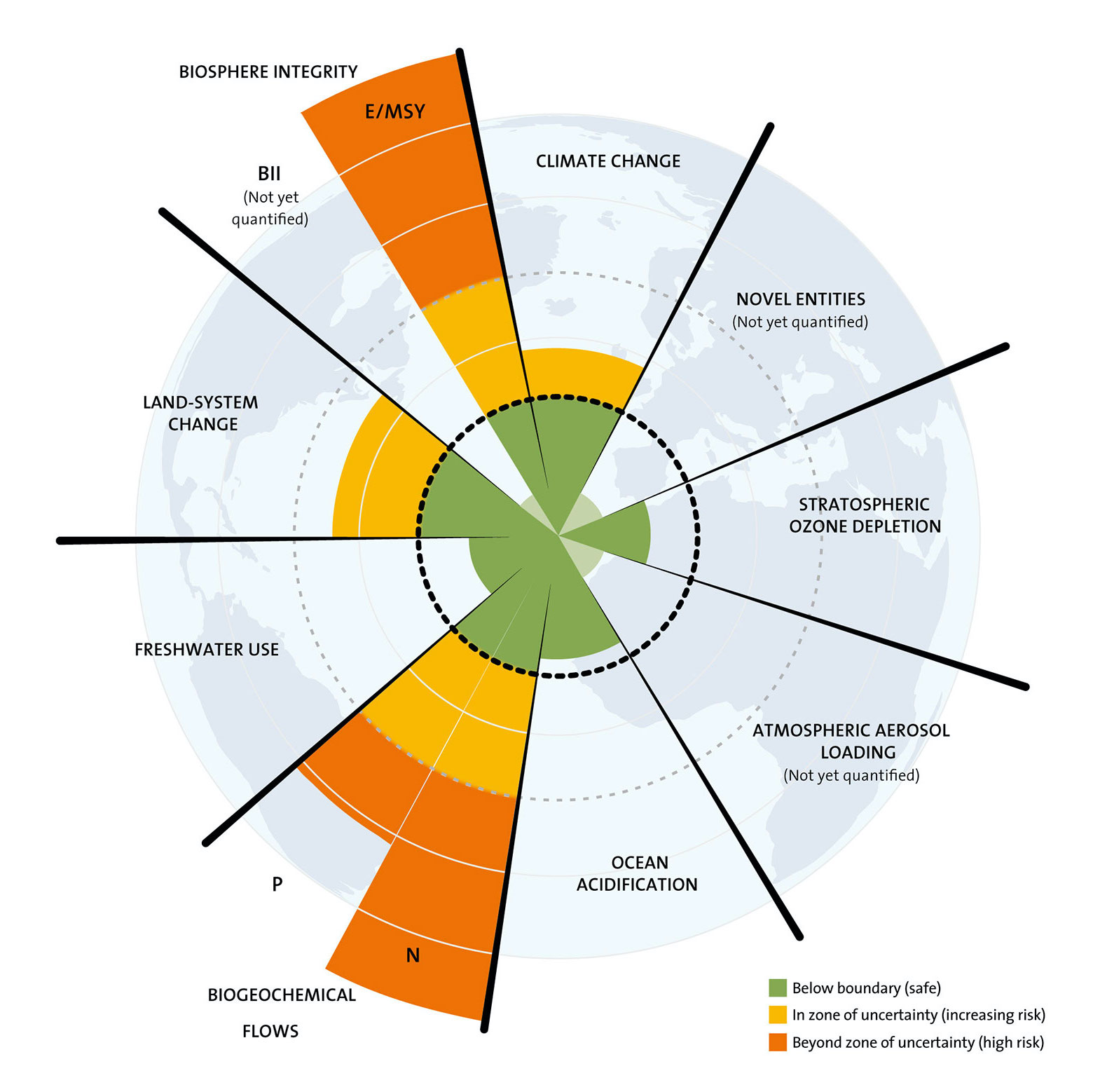}
	\caption{Parameters that determine the state of the Earth System in the Planetary Boundaries framework. The safety zone is shown in green. The yellow colour indicates overshooting beyond the safety zone. Orange colour indicates an overshooting that can lead to irreversible disruption.}
	\label{fig:planetary_boundaries}
\end{figure}

Another important component of the Scientific Utopia concerns the use of the scientific evidence for the understanding of the Earth System \cite{Steffen:2016} and for establishing the extent of the changes and damage that the human action has already inflicted upon it. Indeed, there is an emerging consensus about the notion of the Earth System, the planetary system that comprises the biosphere, including all living biota, and their interactions and feedbacks with the geosphere, the atmosphere, the cryosphere, the hydrosphere and the upper lithosphere. The state of the Earth System is the result of the interaction of many factors: rate of biosphere loss, land system change, global fresh water use, biogeochemical flows (global Nitrogen and Phosphorus cycles), ocean acidification, atmospheric aerosol loading, stratospheric ozone depletion, climate change, chemical pollution, and some others. Determining the optimal operational range for each of these parameters has led to the so-called Planetary Boundaries \cite{Steffen:2015} and to the alarming awareness that the climate changes are not the only evidence available about the destabilizing nature of the human activities. In fact, through the quantification of the above-mentioned parameters we are led to face the worrisome understanding that at least two of these parameters, the biosphere integrity and biogeochemical flows, have well overshoot the safety boundaries as shown in Figure \ref{fig:planetary_boundaries}. The land system use and the climate change parameters are also clearly under stress. 

%-----------------------------------------------------------------------------%

\section{The Physical Model}

The Anthropocene represents a functionally distinct period in the history of the Earth System (ES). The evolution of Earth is now determined mostly by the influence of human activities, rather than by the natural astronomical and geophysical forces that were dominant. There is also evidence of a great acceleration in the changes caused by humans. This is in stark contrast with a remarkable climatic stability of the 11700 years since the end of the last glaciation, the epoch called the Holocene. This implies that the Earth System was in an equilibrium state that has now been disturbed by the increasing human influence in the environment.

\begin{figure}
	\centering
	\includegraphics[width=0.6\columnwidth]{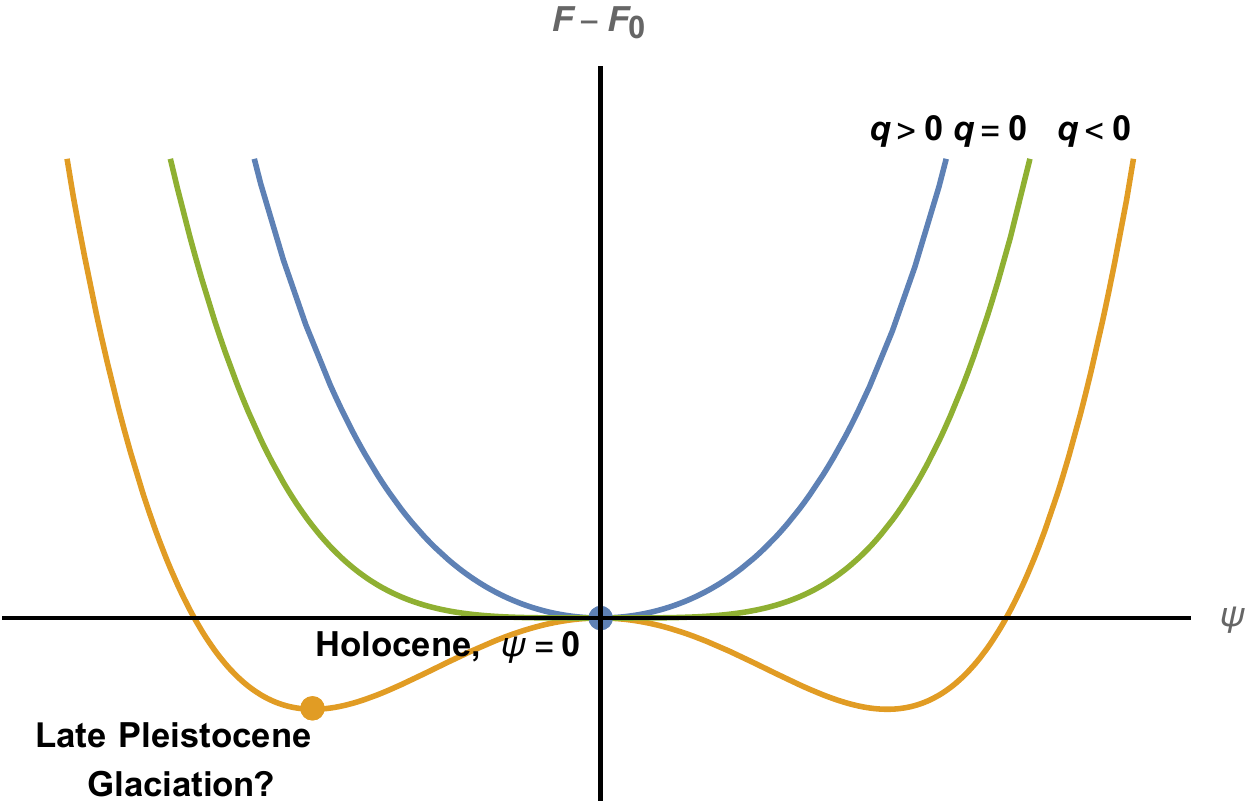}
	\caption{Earth System equilibrium when driven only by natural causes, $q$. All systems tend to minimize their free energy. We can conjecture that the ES has been oscillating between an ``asymmetric'' state ($q < 0$) and a ``symmetric'' state ($q > 0$). The Holocene corresponds to the latter.}
	\label{fig:ESLGT}
\end{figure}

We propose that the transition from the Holocene to the Anthropocene is, in fact, a phase-transition, a qualitative change in the features of the equilibrium state of the Earth System. We have developed a physical model using the Landau-Ginsburg Theory (LGT), a general framework used in all kinds of phase transitions. This kind of description allows one to plot the free energy, $F$, of the system and leads to the identification of the two different phases on each side of the transition and their different equilibrium points, as depicted in Figure \ref{fig:ESLGT}. The key quantity that needs to be defined is the order parameter, $\psi$, which we assume as to be the temperature departure from the Holocene average. Full details of the proposed model are available in Ref. \cite{Bertolami:2018c}.

\begin{figure}
	\centering
	\includegraphics[width=0.6\columnwidth]{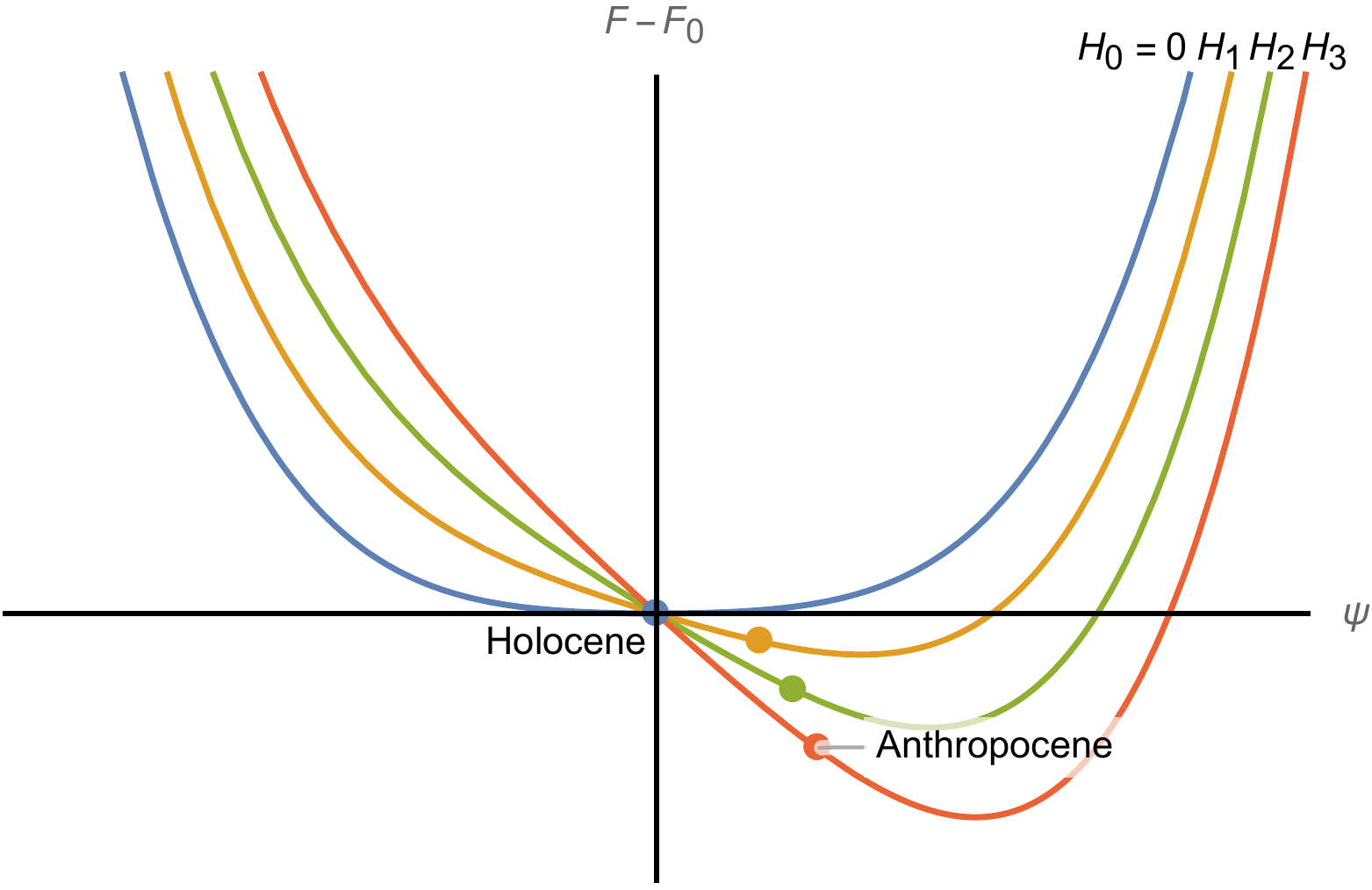}
	\caption{Earth System free energy with increasing values of human activity effects, H. The equilibrium point further departs from the Holocene stability towards higher temperatures. The Anthropocene is a non-equilibrium state where the ES is moving towards a new yet undetermined equilibrium.}
	\label{fig:ESLGTH}
\end{figure}

The Landau-Ginzburg formulation then allows us to study the impact of an additional influence to the Earth system, in this case, the effects of human activities, modelled as an external field, H. The effect is immediately visible in the destabilization of the Holocene conditions, setting the Earth System on a path towards a new yet unknown equilibrium state, as shown in Figure \ref{fig:ESLGTH}. We are also able to compute a susceptibility parameter that shows how the effects of the external field are much greater if the system is close to the phase transition. This is equivalent to saying that there is a tipping point at the phase transition.

The goal of setting up this theoretical framework is that it can be fed with the existing and future data on the conditions of the Earth system and its various subsystems and built a multidimensional predictive model for the Anthropocene. We have already exemplified how this can be carried out for a single parameter, a measurement of biomass depletion, and how the model can be constrained from the data in order to provide numerical predictions.

%-----------------------------------------------------------------------------%

\section{The Earth as a Dynamical System}

Departing from the physical framework mentioned in the previous section, it is possible to build a fully-fledged dynamical systems description of the ES. This will allow for a phase space analysis of the temperature field, to study the set of possible evolution trajectories that the ES can take. This kind of description, even if with incomplete information on the starting conditions, allows for a description of some of the properties and inherently nonlinear behaviour of the ES, without the need to completely solve the underlying differential equations.

That way, we have a mathematical model well grounded on physics that can capture the behaviour of the ES during this transition, including its non-linear and more complex properties. We thus use the Hamiltonian formulation, ubiquitous in most branches of physics, to obtain the evolution equations of the ES and examine its orbits in the phase space, the space of all possible states of the system. It then becomes evident how the increase in human activities progressively deforms the phase space of the ES towards higher temperatures. With this kind of mathematical description of the ES we can start to look for some of its properties, namely the existence and stability of equilibrium points. Indeed, we show how a given evolution of the human influence will affect these points. Even assuming that their effects are bounded, the ES will unquestionably progress towards a new equilibrium away from the Holocene, supporting a Hothouse Earth scenario \cite{Bertolami:2019}.

\begin{figure}
	\centering
	\includegraphics[width=0.8\columnwidth]{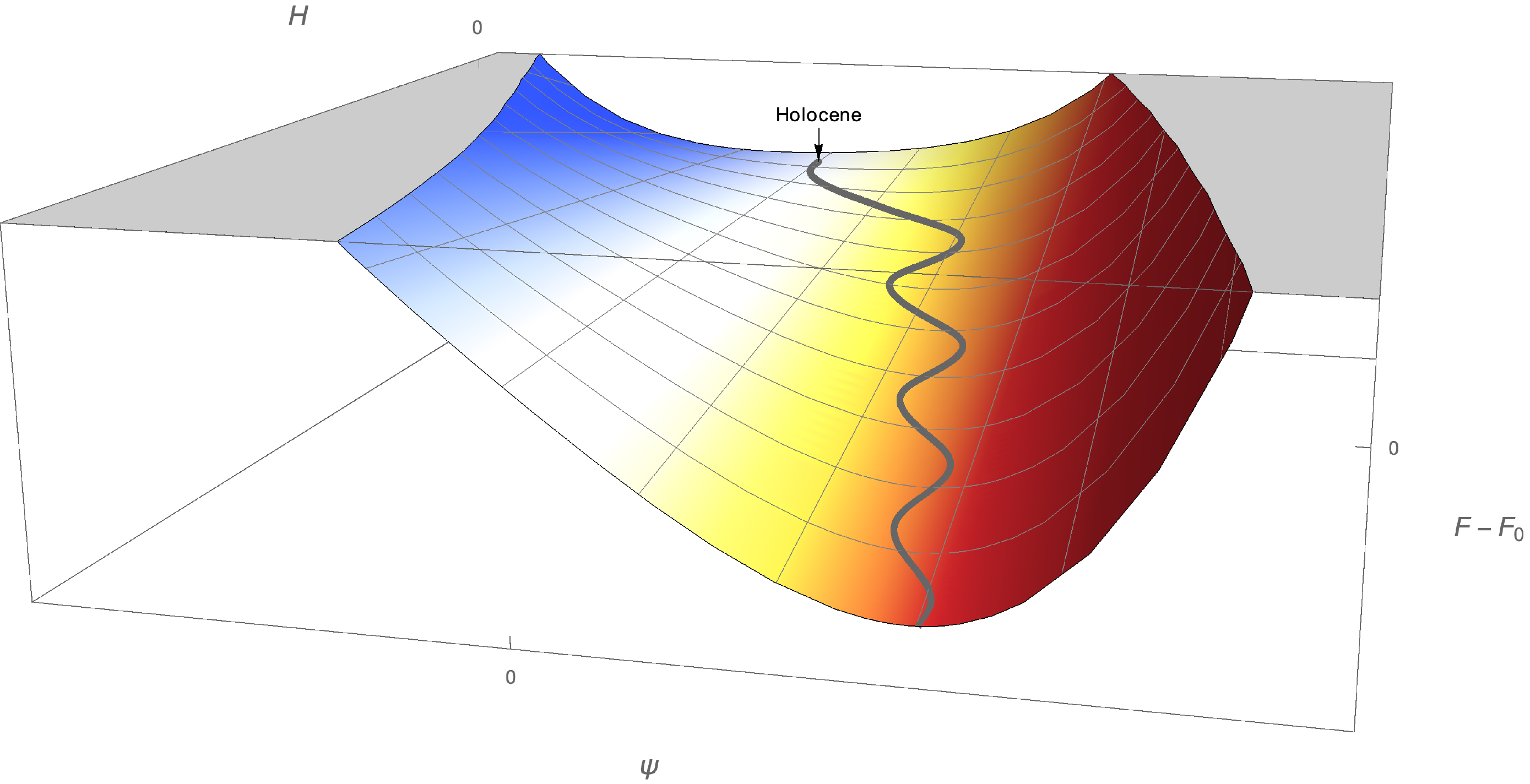}
	\caption{Stability landscape of the ES. $H$ axis measures human activities in terms of the Holocene conditions, that grow with time, $\psi$ axis measures temperature anomaly relative to the Holocene average.}
	\label{fig:stability_landscape_trajectory}
\end{figure}

Using this phase-space description of the ES, we have already shown how it can model the ES trajectories that can lead to a Hothouse Earth scenario, recently discussed in a qualitative fashion by Steffen et al. \cite{Steffen:2018}. The plot in Figure \ref{fig:stability_landscape_trajectory} shows how, for a finite amount of human driven change, the ES enters a new stable equilibrium, i.e. a minimum in the free energy function and, therefore, an attractor of the dynamical system trajectories.

We have also discussed how the human activities field, $H$, can be more accurately modelled, including interactions between its components. The aim is to find a way to quantify the Planetary Boundaries \cite{Steffen:2015} and to check how we can maintain the ES within its Safe Operating Space.

Our formulation does not aim to supersede classical climate models, such as the Budyko-Sellers model or the more recent extensions that also include dynamical elements, but these can be incorporated into our model as one of the ES subsystems, albeit a crucially important one. 

A key issue remains in how to correctly describe the human effects in their multiple components. We stress how these components are not independent and may have complex interactions. We show, at least theoretically, that interactions among these components, most particularly the one associated with the Technosphere, can allow for mitigating strategies in what concerns the inevitable evolution towards a Hothouse Earth, as well as the “engineering” of metastable states where the ES can remain temporarily in equilibrium in a cooler Stabilized Earth scenario.

With a fully dynamical description of the ES in the Anthropocene, the work can now evolve towards a more quantitative analysis along the lines of the Planetary Boundaries variables so that specific predictions can be made. Crucially, we have a way of modelling the different components of human intervention, as well as their interactions.

%-----------------------------------------------------------------------------%

\section{Towards a Physically Motivated Earth\\ Accounting System}

The physical description of the Earth System (ES) in terms of the theory of phase transitions discussed in the previous sections provides a natural accounting framework for measuring the impact of the human drivers once these are broken in terms of its fundamental components, for which the Planetary Boundaries framework provide a natural breakdown.

This suggests some accounting strategies, which can be gauged in terms of population, surface area, or economic output of a given territory. A quota system can then be derived from the planetary boundaries \cite{Barbosa:2019}. Our proposal closely resembles the quota system of Meyer and Newman \cite{Meyer:2018}, but includes interaction between different planetary boundaries. In fact, we have estimated the interaction term between the CO2 concentration in the atmosphere and ocean acidification Planetary Boundaries \cite{Steffen:2018}.

%-----------------------------------------------------------------------------%

\section{Utopias for the Near Future}

Within a human lifetime, we have witnessed an extraordinary intensification of human activities, to the point that they have become ubiquitous. Human activities have affected the entire planet in a wide range of ways. However, transformation is a key feature of Earth and the struggle for survival and underlying element of all life in the planet.

With these premises, the only certainties are the scientific method, humanist values, and the urgency of the problems. Naturally, the consensus about the issues at hand requires a debate that collectively sets the objectives to attain. 

The two examples that follow are among the objectives for humankind that in the next decade bear, in our view, the watermark of utopic goals.

Establishing a legal framework that internalizes the Earth System processes. This would set the ES to be managed and kept in a way akin to a condominium, the Common Home of Humanity \cite{CHH}.

The Great Green Wall in North Africa aiming to contain the advance of the Sahara desert, protect the Sahel, mitigate climate change and improve living conditions for communities of the Sahelian-Saharan countries enrolled in this task (Algeria, Burkina Faso, Benin, Chad, Cape Verde, Djibouti, Egypt, Ethiopia, Gambia, Libya, Mali, Mauritania, Niger, Nigeria, Senegal, Somalia, Sudan, and Tunisia) \cite{TGGW}.

%-----------------------------------------------------------------------------%
%-- Bibliography -------------------------------------------------------------%
%-----------------------------------------------------------------------------%


\begin{thebibliography}{99}

\bibitem{Bertolami:2015} O. Bertolami, A tensão utópica (2015). (\url{http://web.ist.utl.pt/orfeu.bertolami/Utopia_Bertolami.pdf})

\bibitem{Bertolami:2018a} O. Bertolami, Utopia: Utopian and Scientific (2018). (\url{http://web.ist.utl.pt/orfeu.bertolami/Bertolami_Utopia_2018.pdf})

\bibitem{Bertolami:2018b} O. Bertolami, A Humanidade no Antropoceno (2018). (\url{https://forumdemosnet.wordpress.com/2018/12/01/a-humanidade-no-antropoceno/})

\bibitem{Sen:2012} Amartya Sen, Sobre Ética e Economia (Edições Almedina, 2012).

\bibitem{Steffen:2016} W. Steffen, et al., Stratigraphic and Earth System approaches to defining the Anthropocene, Earth’s Future 4, 324 (2016).

\bibitem{Steffen:2015} W. Steffen, et al., Planetary boundaries: Guiding human development on a changing planet, Science 347(6223), 1259855 (2015). doi: 10.1126/science.1259855.

\bibitem{Bertolami:2018c} O. Bertolami and F. Francisco, A physical framework for the earth system, Anthropocene equation and the great acceleration, Global Planet. Change 169, 66–69 (2018). doi: 10.1016/j.gloplacha.2018.07.006

\bibitem{Bertolami:2019} O. Bertolami and F. Francisco, A phase-space description of the Earth System in the Anthropocene (2019). To appear on EPL. arXiv: 1811.05543[physics.ao-ph]

\bibitem{Steffen:2018} W. Steffen, et al., Trajectories of the Earth System in the Anthropocene, Proc. Natl. Acad. Sci. USA, 115(33), 8252–8259 (2018). doi: 10.1073/pnas.1810141115

\bibitem{Barbosa:2019} M. Barbosa, O. Bertolami and F. Francisco, Towards a Physically Motivated Planetary Accounting Framework (2019). Submitted to The Anthropocene Review. arXiv: 1907.10535[physics.ao-ph]
	
\bibitem{Meyer:2018} K. Meyer and P. Newman, The planetary accounting framework: a novel, quota-based approach to understanding the impacts of any scale of human activity in the context of the planetary boundaries, Sustainable Earth 1, 4 (2018).

\bibitem{CHH} Common Home of Humanity. \url{http://www.commonhomeofhumanity.org}

\bibitem{TGGW} The Great Green Wall for the Sahara and the Sahel Initiative (2015). \url{https://www.unccd.int/content/great-green-wall-sahara-and-sahel-initiative}

\end{thebibliography}
\end{document}